\documentclass[sigconf]{acmart}
\AtBeginDocument{%
  \providecommand\BibTeX{{%
    \normalfont B\kern-0.5em{\scshape i\kern-0.25em b}\kern-0.8em\TeX}}}
\usepackage{graphicx,tabularx,subcaption}
\usepackage{booktabs}
\usepackage{multirow}
\usepackage[capitalise,nameinlink,compress]{cleveref}
\usepackage{xcolor}
\usepackage{comment}
\usepackage[english]{babel}
\graphicspath{{images/}}

\copyrightyear{2023} 
\acmYear{2023} 
\setcopyright{acmlicensed}
\acmDOI{10.1145/3581641.3584075}

\acmConference[IUI '23]{28th International Conference on Intelligent User Interfaces}{March 27--31, 2023}{Sydney, NSW, Australia}
\acmBooktitle{28th International Conference on Intelligent User Interfaces (IUI '23), March 27--31, 2023, Sydney, NSW, Australia}
\acmPrice{15.00}
\acmISBN{979-8-4007-0106-1/23/03}

\begin{document}

\title[Directive Explanations for Monitoring the Risk of Diabetes Onset]{Directive Explanations for Monitoring the Risk of Diabetes Onset: Introducing Directive Data-Centric Explanations and Combinations to Support What-If Explorations}

\author{Aditya Bhattacharya}
\orcid{0000-0003-2740-039X}
\email{aditya.bhattacharya@kuleuven.be}
\affiliation{%
  \institution{KU Leuven}
  \city{Leuven}
  \country{Belgium}
}
\author{Jeroen Ooge}
\orcid{0000-0001-9820-7656}
\email{jeroen.ooge@kuleuven.be}
\affiliation{%
  \institution{KU Leuven}
  \city{Leuven}
  \country{Belgium}
}
\author{Gregor Stiglic}
\orcid{0000-0002-0183-8679}
\email{gregor.stiglic@um.si}
\affiliation{%
  \institution{University of Maribor}
  \city{Maribor}
  \country{Slovenia}
}

\author{Katrien Verbert}
\orcid{0000-0001-6699-7710}
\email{katrien.verbert@kuleuven.be}
\affiliation{%
  \institution{KU Leuven}
  \city{Leuven}
  \country{Belgium}
}

\renewcommand{\shortauthors}{Bhattacharya, et al.}

\begin{abstract}
Explainable artificial intelligence is increasingly used in machine learning (ML) based decision-making systems in healthcare. However, little research has compared the utility of different explanation methods in guiding healthcare experts for patient care. Moreover, it is unclear how useful, understandable, actionable and trustworthy these methods are for healthcare experts, as they often require technical ML knowledge. This paper presents an explanation dashboard that predicts the risk of diabetes onset and explains those predictions with data-centric, feature-importance, and example-based explanations. We designed an interactive dashboard to assist healthcare experts, such as nurses and physicians, in monitoring the risk of diabetes onset and recommending measures to minimize risk. We conducted a qualitative study with 11 healthcare experts and a mixed-methods study with 45 healthcare experts and 51 diabetic patients to compare the different explanation methods in our dashboard in terms of understandability, usefulness, actionability, and trust. Results indicate that our participants preferred our representation of data-centric explanations that provide local explanations with a global overview over other methods. Therefore, this paper highlights the importance of visually directive data-centric explanation method for assisting healthcare experts to gain actionable insights from patient health records. Furthermore, we share our design implications for tailoring the visual representation of different explanation methods for healthcare experts.
\end{abstract}




\maketitle

\section{Introduction}
Machine Learning (ML) based systems have been increasingly adopted in healthcare over the past few decades, in applications ranging from surgical robots to automated medical diagnostics \cite{1}. Especially for screening and monitoring of diseases such as type-2 diabetes, ML models have proven to be significant \cite{20, 21}. However, most of these algorithms are ``black-boxes'' because the reasoning behind their predictions is unclear \cite{2}. Moreover, the growing concern of bias, lack of fairness, and inaccurate model prediction have limited the adoption of ML more recently \cite{6}. 

Consequently, \textit{explainable artificial intelligence} (XAI) has gained a lot of focus from ML practitioners as XAI methods facilitate the interpretation and understanding of complex algorithms, thereby increasing the transparency and trust of such black-box models \cite{48, 1, Miller2017}. In healthcare, XAI empowers medical experts to make data-driven decisions using ML, resulting in a higher quality of medical services \cite{3} and can impact its trust and reliance \cite{57, 49}.

Existing XAI methods \cite{7, 8, 9, 5} are predominantly designed for ML practitioners instead of \textit{non-expert users} \cite{29}, who might be specialized in a particular application domain but lack ML knowledge \cite{10}. Yet, the effectiveness of these explanation methods has not been fully analyzed due to the lack of user studies with non-expert users \cite{49, 50}. This gap highlights the necessity for analyzing and comparing explanation methods with healthcare professionals (HCPs) such as nurses and physicians \cite{51} as it is unclear how useful, understandable, actionable, and trustworthy these methods are for them.

Moreover, non-expert users need help to understand how to obtain a favorable outcome \cite{13, 14, 15}. This emphasizes the need to make explanations \textit{directive}, i.e. guiding the users to take action for achieving their desired outcome \cite{13}.  Additionally, instead of \textit{static} explanations, non-expert users have considered \textit{interactive} explanations essential to support understanding and interpretation \cite{52, 53, Kulesza2015}. Therefore, \textit{visually directive explanations} should enable non-experts not only to understand \textit{why} a certain outcome is predicted but also to guide them in the process of finding \textit{how} to obtain their desired outcome without any intervention from ML experts \cite{16, 17, 18, 19}. 

We designed a prototypical dashboard that predicts patient's risk of diabetes onset using visually directive explanations based on different explanation methods, including data-centric approaches \cite{47}, feature importance \cite{32}, and example-based methods \cite{32}. We aimed to support nurses and physicians in screening patients with undiagnosed type-2 diabetes, monitoring their conditions, and suggesting actions to control their risk of diabetes onset. 

We also obtained the perspective of diabetic patients during the evaluation process as they are well aware of the risk factors of type-2 diabetes. Furthermore, some of them were recently in the pre-diabetes phase and all of them are actively in contact with their HCP. Thus, we analyzed their motivation to use such a dashboard. 

This paper probes into the following research questions: 

\begin{description}
\item[RQ1.] In what ways do patients and HCPs find our visually directive explanation dashboard useful for monitoring and evaluating the risk of diabetes onset?
\item[RQ2.] In what ways do HCPs and patients perceive data-centric, model-centric, and example-based visually directive explanations in terms of usefulness, understandability, and trustworthiness in the context of healthcare?
\item[RQ3.] In what ways do visually directive explanations facilitate patients and HCPs to take action for improving patient conditions?
\end{description}

We explored these questions through a two-phased study: first, a qualitative study on a low-fidelity click-through prototype involving 11 HCPs; and second, a mixed-methods online study for the evaluation of a high-fidelity web application prototype involving 51 patients and 45 HCPs. We analyzed the effectiveness of our different visual explanation methods and compared them in terms of understandability, usefulness, actionability, and trust \cite{Hoffman2018, Sokol_2020}. Our results show that our dashboard provided actionable insights to HCPs about patient health by helping them to identify important risk factors and showcase how critical the patients’ conditions are. Additionally, it helped patients to self-monitor and analyze their health conditions.

This paper presents three primary research contributions. First, we present our visually directive data-centric explanation methods that are aimed to provide local explanations of the predicted risk for individual patients with a global overview of risk factors for the entire patient population. Whereas it has been shown that non-expert users prefer local explanations that justify a single decision \cite{29}, it has also been argued that these explanations rarely provide sufficient insight into the reasoning of models and the explanatory depth that non-experts require to accept and trust the decision-making of the model \cite{dazeley2021levels}. To address this challenge, we present an approach that combines perspectives of both local and global explanation methods \cite{30} to provide more insight into both the model predictions and the data for non-expert users. 
Second, we present the design of a dashboard that combines different explanation methods based on an iterative user-centered research process. Third, based on observations of our user-centered design process and an elaborate user study, we present design implications for tailoring explanations for healthcare experts. We observed that our participants had a higher preference for our representation of data-centric explanations over other methods as they found them more informative. We also observed that participants combined multiple explanation methods, particularly for recommending actions to minimize the risk and interpreting the rationale behind the predicted risk. Based on these observations, we present design implications for tailoring directive explanations for healthcare experts.




\section{Background and related work}
Designing visually explainable Decision Support System (DSS) in healthcare considering different types of explanations is an active area of research in XAI \cite{22}. To contextualize our research, we first review recent research findings in the domain of visually interactive DSS in healthcare and then investigate XAI methods that provide visual explanations to end-users.

\subsection{Visually Interactive DSS in Healthcare}
In healthcare, using DSSs built on the domain knowledge of medical experts has a long history \cite{23}. Usually, such systems are rule-based logical systems developed on pre-defined rules supplied by medical experts \cite{24}. Despite the explainability offered by such systems, there are many challenges such as poor user experience and scarcity of involvement of medical experts in forming the knowledge base \cite{25,26}. 

To overcome these challenges, modern DSSs in healthcare use ML and data-driven techniques to learn patterns from historical data and apply visualizations to facilitate prescriptive insights for medical practitioners \cite{26,27, Holzinger_2017}. ML-based DSSs are being increasingly used in healthcare for the early detection of health conditions such as undiagnosed type-2 diabetes mellitus \cite{21}. Despite the success of such systems, the lack of transparency of advanced ML algorithms has increased the need for human-friendly explainable DSS in healthcare \cite{28}. To mitigate these challenges, interactive interfaces have proven to improve the understanding of non-expert users \cite{31}. 

Moreover, many researchers have found additional benefits in applying XAI for clinical DSSs such as the mitigation of cognitive bias \cite{35}. Our research work focuses on providing an explainable DSS which is interactive and personalized to meet the needs of the medical experts involved in the progressive monitoring of the risk of diabetes onset for patients.

\subsection{Exploration in Visual Explanations}
Harmonizing XAI techniques with \textit{visual explanations} enable non-expert users to gain appropriate trust in the outcome of ML systems \cite{12}. Recent works also suggest that exploration and contextualization of explanation methods can enhance the satisfaction and interpretability of non-expert users \cite{29}.

Model-agnostic post-hoc explanation techniques \cite{Langer_2021, Sokol_2020, 3} explain black-box ML models without having any intrinsic information about the inner working of the algorithm, i.e. knowledge about inner parameters or hyper-parameters of the model.  Most common model-agnostic local explanation methods like LIME \cite{7}, and SHAP \cite{8} are feature-importance-based methods that identify the most impactful features contributing to the model’s prediction \cite{32}. 

However, more recently, due to the failure of ML models trained on biased, inconsistent and poor-quality data, the ML research community is exploring data-centric approaches \cite{47, 33}. Examples of data-centric approaches are summarizing individual data instances (using common statistical methods like mean, mode, and variance), visualizing the data distribution to compare feature values of an instance to those across the remaining dataset and observing changes in model predictions through \textit{what-if analysis} \cite{34, 35, 36, 37}. Additionally, data-centric explanations include data-driven rule-based approaches that are adopted commonly in medical DSS for assisting health experts \cite{22, 23, 25, 28, 37}. 

Additionally, researchers have used counterfactuals to provide recommendations for health-related changes \cite{32, 12, 35}. Adadi and Berrada \cite{32} have defined counterfactual explanations as \textit{example-based} methods that provide minimum conditions required to obtain an alternate decision. Although counterfactuals provide useful model-agnostic post-hoc explanations, examples generated by counterfactual algorithms can be practically infeasible, contradictory, or uncontrolled, thereby indicating a need for actionable recourse \cite{40, 15}. For instance, to obtain a lower risk of diabetes, counterfactual algorithms can indicate patients to reduce their age by 30 years or alter their gender, which is practically infeasible. Yet, visually interactive counterfactuals hold great potential to produce actionable insights \cite{34}. Thus, there is an opportunity to explore a better representation of such explanation methods for achieving actionable recourse. 

Moreover, as established by Bove et al. \cite{29}, exploring explainable interfaces is considered essential for the interpretation and satisfaction of end-users. The same notion is adopted in our work for considering different types of visual explanation methods.

\section{Material and Methods}
This section presents our visually directive explanation dashboard and our user-centric methodology for the design and evaluation of our prototypical dashboard. The ethical approval for our research was granted by the ethical committee of KU Leuven with the number G-2019-09-1742.

\begin{figure*}
  \centering
  \frame{\includegraphics[width=0.90\linewidth]{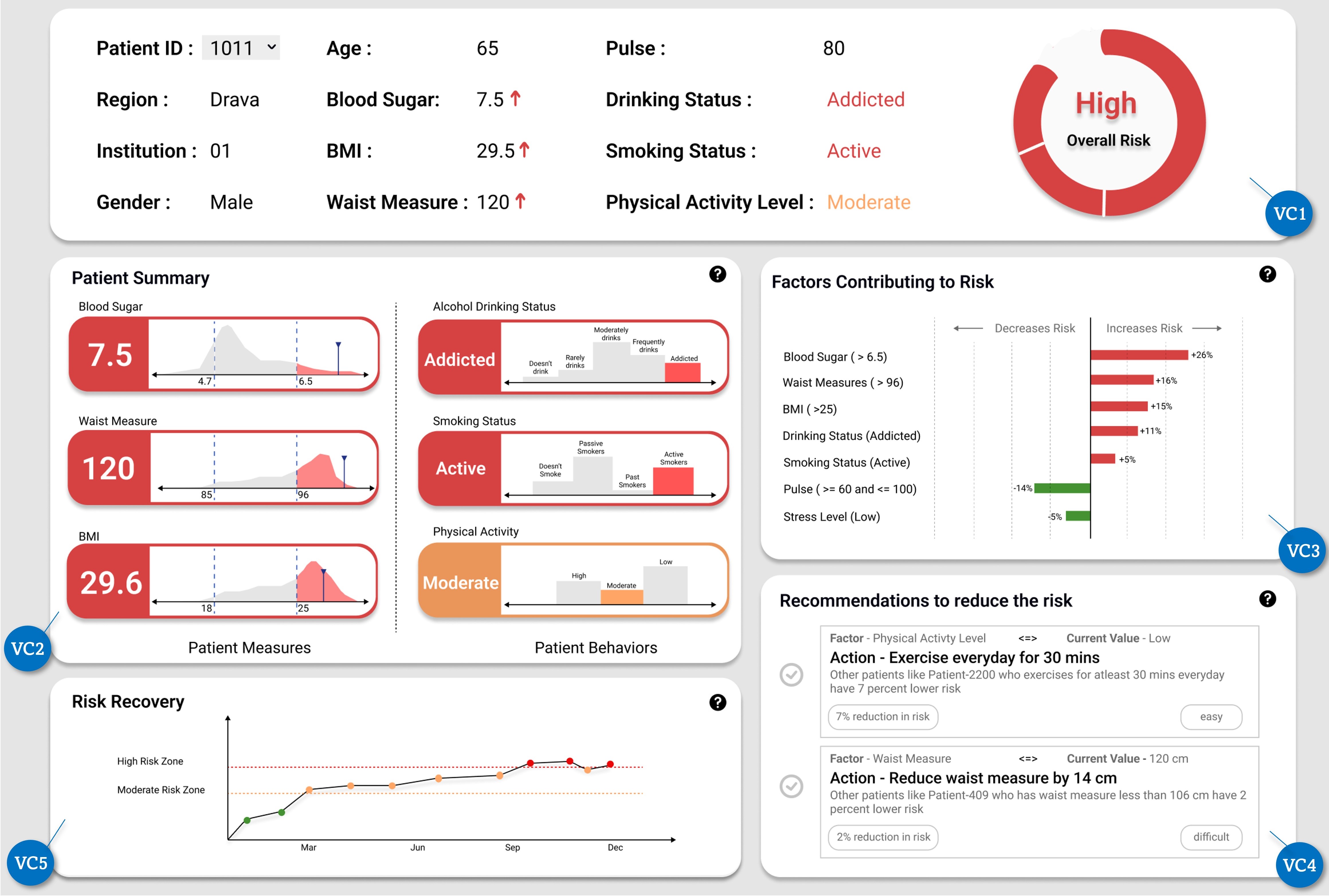}}
  \caption{Dashboard design of our click-through prototype. Visual explanations are provided using: Patient information with the risk prediction chart (VC1), Patient Summary (VC2), Factors Contributing to Risk (VC3), Recommendations to reduce risk (VC4), Risk recovery (VC5).}
  \Description[Dashboard design of our click-through prototype]{Dashboard design of our click-through prototype}
  \label{fig:low_fi}
\end{figure*}

\subsection{Visually Directive Explanation Dashboard} 
Our explanation dashboard is designed to assist HCPs in monitoring and screening the risk of diabetes onset for patients. We enabled users to explore different kinds of visual explanations for interpreting the model prediction.
\paragraph{ML Model:} We used a logistic regression algorithm from Python scikit-learn module to train a classifier on our diabetes healthcare dataset. We achieved a training accuracy of 93\% and a test accuracy of 91\% (no overfitting effect). The test accuracy is considered as overall model accuracy. The complete technical approach of data processing, model training, tuning and evaluation was conducted in Python
\paragraph{Dataset:} Our ML model was trained on patient’s electronic health records of comprehensive medical examinations conducted from 5 Slovenian primary care institutions \cite{21}. The health records include information about patients such as blood glucose, waist circumference measure, BMI, age, gender, etc., along with patient behaviors collected from the Finnish Diabetes Risk Score questionnaire (FINDRISC) to predict the risk of diabetes. 

Our dashboard explained the predicted risk for an individual patient along with an overview of the health statistics of the entire patient population in the medical records. Thus, our approach provides \textit{local explanations} with a \textit{global perspective} to HCPs \cite{30}.

\subsubsection{\textbf{User requirements}}
We conducted a preliminary one-hour focus group with four healthcare workers with active experience in patient-care from Community Healthcare Centre dr. Adolf Drolc Maribor, Slovenia to understand their needs and challenges while monitoring the risk of diabetes in patients. We first demonstrated a dashboard based on the work of Rojo et al. \cite{55}, AHMoSe, that explains the predicted risk of diabetes using SHAP values. Then, we did a co-design session with our participants to tailor the AHMose interface for meeting their requirements. The session was recorded and transcribed for analyzing their feedback and responses.

From this exploratory study, we identified their struggles in monitoring patients for the onset of diabetes. We also learned that the visualizations for SHAP based explanations in AHMose were too complex and detailed. Our participants wanted an easy way to find only health factors which plays a vital role in elevating the risk of diabetes for patients, instead of looking into all the features considered by the model for making predictions. We also learned from the co-design activity that the healthcare workers preferred simpler line-chart, bar-chart and textual representations of the data that could also be used to communicate with patients. 


Additionally, we asked our about their specific needs in this study. We summarize their responses into the following user requirements that our explanation system should meet:
\begin{description}
\item[1. An interface for monitoring patients] – HCPs wanted a visual interface to quickly observe the medical records of patients as it is inconvenient and time-consuming to examine multiple medical reports to assess patient health conditions. Additionally, HCPs wanted to analyze how a specific patient is doing compared to other patients. 
\item[2. Suggest actions to minimize the predicted risk] - HCPs wanted to use an interactive interface to suggest actions to patients to minimize their risk of diabetes onset. 
\item[3. Increase patient awareness] - HCPs wanted the interface to show patients how critical their conditions are for creating more conscious awareness and motivating them to follow the prescribed suggestions sincerely.
\end{description}

Then we designed our tailored explanation dashboard which supported the following tasks for meeting the user requirements:

\begin{description}
\item[T1: Monitor the risk of diabetes for patients] – Understand the current risk of diabetes onset for patients and identify if the patient’s condition is more critical or not for deciding the amount of attention needed for the patient.
\item[T2: Propose actions to minimize the predicted risk] – Suggest actions to patients to reduce the risk of diabetes or to keep it under control.
\item[T3: Interpret the rationale behind the predicted risk] – Understand and explain the system’s logic for the estimated risk of diabetes by identifying the health variables and their range of values that can increase or decrease the risk of diabetes.
\item[T4: Compare the current patient with other patients] – By comparing the health measures of the current patient with other patients, get an indication about a specific patient’s situation as compared to other patients to decide the amount of attention needed. 
\end{description}

Task \textbf{T1} aims to meet the first requirement, \textbf{T2} aims to meet the second requirement and \textbf{T3} and \textbf{T4} aims to meet the third requirement.

\subsubsection{\textbf{XAI techniques and visual components}}

\label{sec:3_4_xai_techniques} When we analyzed our user requirements, we found that these requirements are aligned with the \textit{explanation goals} presented by Wang et al. \cite{35}. Therefore, the choice of our explanation methods should facilitate learning by enabling our users to filter a small set of factors to make their observations simpler and provide them with the ability to predict and control future phenomena by generalizing these observations into a conceptual model. Wang et al. \cite{35} also proposed \textit{XAI elements} that can meet these explanation goals and recommended visualizations that can be used to present these XAI elements.
Moreover, we considered \textit{model-agnostic} \textit{local} explanation methods for explaining the predicted risk for individual patients irrespective of the ML algorithm used. We further designed visual components as illustrated in \Cref{fig:low_fi} for the following three types of explanation methods that meet our explanation goals:

\textbf{Feature Importance explanation} – As the dashboard aimed to direct HCPs towards suggesting actions to patients for minimizing the risk, \textit{feature-importance explanations} enabled them to identify the most influential risk factors according to the prediction model. However, Szymanski et al. \cite{39} have shown that the representation of feature-importance explanations can impact the understandability and usefulness of this method. Additionally, from our preliminary focus group session, we observed that our participants did not understand simplified SHAP-based feature-importance explanations in the AHMoSe dashboard \cite{55}. 

Our representation of directive feature-importance explanations presented in \textit{Factors Contributing to Risk} (\textbf{VC3}) included only the \textit{actionable health variables} with a percentage measure that indicated how specific features influenced the prediction. We define \textit{actionable health variables} as variables that can be controlled by the patient such as BMI, waist circumference, physical activity level. We considered other factors that are infeasible for the patient to alter such as age, gender, and geographical region as \textit{non-actionable health variables}.

The feature-importance scores are calculated using the SHAP Python module. Factors that can increase risk are displayed in red, while those which can decrease risk are displayed in green. Subtle explanations are provided beside each health variable by comparing the feature value with the recommended range to indicate why it can increase or decrease risk.

\textbf{Data-Centric explanations} – Data-centric explanations included in our dashboard aimed to explain why a certain health variable is increasing or decreasing the risk of diabetes without focusing on what the model considered important. \textit{Patient information with risk prediction chart} (\textbf{VC1}), \textit{Patient Summary} (\textbf{VC2}), and \textit{Risk Recovery} (\textbf{VC5}) provided data-centric explanations in our dashboard. 

\textbf{VC1} gave a textual overview of the health information of a particular patient with a doughnut chart showing the predicted risk of diabetes onset. To enable HCPs to observe the variance in the predicted risk between different patients we categorized the probability score (\textit{prob}) generated by our ML classifier into three levels: \textit{High (prob > 0.75)}, \textit{Moderate (0.5 $\leq$ prob $\leq$ 0.75)} and \textit{Low (prob < 0.5)}. The abstracted level is displayed in the center of the doughnut chart, and the risk percentage (\textit{prob * 100\%}) as a tool-tip. Moreover, we used consistent color coding across the different visual components (\textit{red} for \textit{high}, \textit{orange} for \textit{moderate}, and \textit{green} for \textit{low} risk). We also provided subtle indicators like colored arrows for numerical feature variables and colored text for categorical feature variables, along with necessary tool-tips to indicate if the specific health variable is within the recommended range or not. However, these visual indicators are only provided for \textit{actionable health variables}.

\textbf{VC2} showed the value of each actionable health variable of a patient along with data-distribution charts, considering the entire patient population in the records. This component enabled HCPs to have a quick overview of a specific patient as compared to other patients and observe common patterns across all patients using the data-distribution charts. The health variables used in the data-distribution charts are further segregated as patient \textit{measures}, which is visually represented with area charts as these are continuous variables, and \textit{behaviors} which is visually represented with bar charts as these are categorical variables. \textbf{VC2} also showed the recommended ranges for the patient measures as configured by the health experts. It can be used to observe the health status of a patient in comparison to other patients. The data-distribution zone where the current health measure lies in the distribution chart is also color-coded with our consistent color-coding convention.

\textbf{VC5} enabled progressive monitoring of the predicted risk of diabetes considering the historical patient records. It was designed to allow users to identify if the patient’s condition is improving or not over a period of time. 


\textbf{Counterfactual explanations} – \textit{Recommendations to reduce risk }
 (\textbf{VC4}) suggested actions using counterfactual explanations generated by the DiCE framework \cite{56} that patients can take to reduce the predicted risk of diabetes. To mitigate the drawbacks of the counterfactual algorithm implemented in the DiCE framework \cite{56}, we considered generating counterfactuals for only actionable health variables instead of non-actionable variables. 
 
 We also added data-driven boundary conditions so that counterfactuals with absurd alterations are avoided. Furthermore, the recommendations are presented as textual statements instead of discrete numbers for easier interpretation. We considered having these textual recommendations pre-configured for the patient behaviors that are categorical features of the model. For example, instead of suggesting the patient to increase their physical activity level from low to moderate, the visual recommends they exercise daily for 30 minutes.

\textbf{VC4} also included an indication of feasibility (\textit{easy} or \textit{difficult}) and an estimated measure of risk reduction using \textit{sensitivity analysis} \cite{32, 18, 37} to compare between different recommended tasks. For continuous numerical features of the model, feasibility is measured by calculating the percentage change between recommended measure value and the current health measure. If the percentage change is within ${\pm}$10\%, feasibility is labeled as \textit{easy}. Otherwise, it is considered as \textit{difficult}. For categorical model features, ordinal integer encoding is done based on how the specific value can increase or decrease the risk of diabetes, and feasibility is considered based on the encoded ordinal values. For instance, \textit{physical activity level} is a categorical variable having three possible values: \textit{low}, \textit{moderate}, and \textit{high}. \textit{Low} physical activity can increase the risk, and hence the corresponding ordinal value of 1 is assigned to it. If the value is \textit{moderate}, an ordinal value of 2 is assigned, and if it is \textit{high} an ordinal value of 3 is assigned. Any change to immediate ordinal value is considered \textit{easy}. For instance, a change from \textit{low} to \textit{moderate} is considered \textit{easy}. But otherwise, it is considered as \textit{difficult}. With this approach, we aimed to make counterfactual explanations more useful and actionable in a controlled way.      

\subsection{Evaluation Process}
We evaluated our prototype in a two-phased user study. First, a qualitative study for evaluating a low-fidelity click-through prototype was conducted through individual interviews involving 11 HCPs. The goal of this study was to get early feedback on how our primary users (HCPs like nurses and physicians) perceive the dashboard and whether the user requirements are being met or not. 

Second, we conducted a mixed-methods study involving 45 HCPs and 51 diabetic patients through online survey questionnaires. The goal of this study was to evaluate the effectiveness of the dashboard in meeting the needs of HCPs and patients. We also compared the different explanation methods represented by our five visual components in terms of understandability, usefulness, actionability, and trustworthiness. The patient’s perspective was also collected from this study as our dashboard would directly or indirectly impact them.


The main similarity between these two user studies is that our participants were given similar task-based questions about the four supported tasks (\textbf{T1}, \textbf{T2}, \textbf{T3}, \textbf{T4}) by our prototype in both studies. Regarding the differences, our first study involved only HCPs, and we recorded the measures of our qualitative evaluation of the visual components through 1:1 interviews. In our second study, we involved a larger pool of participants (including patients) to evaluate our high-fidelity prototype. We recorded our measures of the slightly modified visual components (\Cref{fig:hi_fi}) using a combination of participant-reported qualitative responses and self-reported Likert scale responses.

\section{Evaluation and Analysis of Low-fidelity Prototype}
For our first study, we designed a click-through prototype in Figma \cite{figmaUrl} in multiple iterations.  \Cref{fig:low_fi} illustrates the final design of our low-fidelity click-through prototype.

\begin{table}[h]
\caption{Participants' information for the qualitative evaluation of the low-fidelity prototype.}
\label{tab:participants_study1}
 \scalebox{.75}{
\begin{tabular}{ll}
\toprule
& \textbf{Participant distribution}                                                 
\\ \midrule
Gender & \begin{tabular}[c]{@{}l@{}}
3 : Male \\
8 : Female
\end{tabular}                                                                        
\\ \midrule
Age group & \begin{tabular}[c]{@{}l@{}}
8 : (21-30) years\\ 
1 : (31 - 40) years\\
1: (41 - 50) years\\ 
1: (51-60) years
\end{tabular}                                                                        
\\ \midrule
Highest education level & \begin{tabular}[c]{@{}l@{}}
11 : Master's degree
\end{tabular}                                                                        
\\ \midrule
Experience in patient care  & \begin{tabular}[c]{@{}l@{}}
10 : > 1-year experience \\
in direct patient interaction  \\ 
1: Indirect patient interaction \\
through training of nurses
\end{tabular} \\ 
\bottomrule
\end{tabular}}
\end{table}

\subsection{Participants}

We conducted a qualitative study involving 11 HCPs (male: 3, female: 8) from  the University of Maribor to evaluate our low-fidelity prototype. We recruited participants who had backgrounds in nursing and patient care.  Participants recruited for this study belonged to the same country as the participants of our focus group discussion. But they belonged to two different institutions. \Cref{tab:participants_study1} presents the demographic information of our participants. Our participants reported having experience in diverse specialization areas of healthcare like surgical nursing, interventional radiology, plastic and reconstructive surgery, pediatrics, orthopedics, preventive care, and others. Only one participant had explicit experience in preventive care of diabetes patients. Ten participants had at least one year of experience in looking after patients and having frequent direct interactions with patients. One of the participants was only involved in the training of nurses and did not have any active interaction with patients.

\subsection{Procedure}

The study was conducted through semi-structured individual interviews that lasted between 45-70 minutes, that were recorded and transcribed. During each interview, we introduced our click-through prototype to our participants with brief contextual information. 

Our participants were first asked to explore the prototype, and then asked questions based on the four tasks (\textbf{T1, T2, T3, T4}) supported by the prototype. Each question was followed by necessary follow-up questions about each visual component to understand how effective each component was for performing the given tasks.

We performed a thematic analysis considering the 6-phase-method from Braun and Clarke \cite{42} on the qualitative data. We first reviewed the transcripts of the recorded interviews. Then, we created a list of initial codes from the data. In multiple iterations, we grouped the identified codes into potential themes. After reviewing the preliminary set of themes, we formed a definitive set of themes and accordingly grouped them to analyze how each participant succeeded or failed to answer the tasks-based questions.

To preserve the anonymity of the participants while presenting the results of this study, we refer to the participant as P(N), where N is a particular participant from 1 to 11. We only made necessary grammatical corrections to the participants’ quotes when presenting the results.

\begin{table*}[h]
\caption{Observation from our first user study. The task-based questions are: \textbf{T1(Q1)}: \textit{What is the overall risk for the patients?} \textbf{T1(Q2)}: \textit{Is the condition improving?} \textbf{T2(Q1)}: \textit{What actions can you suggest to the patient to reduce the risk?} \textbf{T3(Q1)}: \textit{Can you explain why the system is showing a high risk of diabetes?} \textbf{T3(Q2)}: \textit{Does the system allow you to see what happens if the blood sugar is less than six instead the current value (7.5)?} \textbf{T4(Q1)}:\textit{ What can you tell about the health variables of the patient as compared to other patients?} \checkmark denotes that the participant was successful in answering the questions, and $\times$ denotes they failed to answer. Visual component(s) used by them to successfully respond to the task-based questions are mentioned in brackets. } 
\label{tab:qual_study_1}
\scalebox{.7}{
\begin{tabular}{c|cc|c|cc|c|}
\cline{2-7}  & \multicolumn{2}{c|}{\textbf{T1}}                               & \textbf{T2} & \multicolumn{2}{c|}{\textbf{T3}}                                                                                                                                                                                                            & \textbf{T4}                                                                                             \\ \cline{2-7} 
                          & \multicolumn{1}{c|}{\textbf{Q1}} & \textbf{Q2} & \textbf{Q1} & \multicolumn{1}{c|}{\textbf{Q1}} & \textbf{Q2} & \textbf{Q1} \\ \hline
\multicolumn{1}{|c|}{P1}  & \multicolumn{1}{c|}{\checkmark  \{VC1\}}                                    & \checkmark \{VC1, VC2, VC3, VC5\}      & \checkmark  \{VC4\}                                                         & \multicolumn{1}{c|}{\checkmark \{VC2\}}                                                                            & \checkmark \{VC1, VC2\}                                                                                                     & $\times$                                                                                              \\ \hline
\multicolumn{1}{|c|}{P2}  & \multicolumn{1}{c|}{\checkmark  \{VC2, VC1\}}                               & $\times$                               & $\times$                                                                   & \multicolumn{1}{c|}{\checkmark  \{VC2\}}                                                                            & $\times$                                                                                                                    & \checkmark \{VC2\}                                                                                    \\ \hline
\multicolumn{1}{|c|}{P3}  & \multicolumn{1}{c|}{\checkmark \{VC2\}}                                    & \checkmark \{VC2\}                     & \checkmark \{VC1, VC2\}                                                    & \multicolumn{1}{c|}{\checkmark \{VC1, VC2, VC4\}}                                                                  & $\times$                                                                                                                    & \checkmark \{VC2\}                                                                                    \\ \hline
\multicolumn{1}{|c|}{P4}  & \multicolumn{1}{c|}{\checkmark \{VC1, VC2, VC3, VC4, VC5\}}                & \checkmark \{VC1, VC2, VC3, VC4, VC5\} & \checkmark \{VC1, VC2, VC3\}                                               & \multicolumn{1}{c|}{\checkmark \{VC2, VC3\}}                                                                       & $\times$                                                                                                                    & $\times$                                                                                              \\ \hline
\multicolumn{1}{|c|}{P5}  & \multicolumn{1}{c|}{\checkmark \{VC1, VC2\}}                               & \checkmark \{VC5\}                     & \checkmark \{VC2, VC3, VC4\}                                               & \multicolumn{1}{c|}{\checkmark \{VC3\}}                                                                            & $\times$                                                                                                                    & $\times$                                                                                              \\ \hline
\multicolumn{1}{|c|}{P6}  & \multicolumn{1}{c|}{\checkmark \{VC1, VC2\}}                               & $\times$                               & \checkmark \{VC1, VC2, VC3, VC4\}                                          & \multicolumn{1}{c|}{\checkmark \{VC3\}}                                                                            & \checkmark \{VC2\}                                                                                                          & \checkmark \{VC1, VC2\}                                                                               \\ \hline
\multicolumn{1}{|c|}{P7}  & \multicolumn{1}{c|}{\checkmark \{VC1\}}                                    & \checkmark \{VC5\}                     & \checkmark \{VC4\}                                                         & \multicolumn{1}{c|}{\checkmark \{VC1, VC2, VC3, VC4, VC5\}}                                                        & \checkmark \{VC2\}                                                                                                          & $\times$                                                                                              \\ \hline
\multicolumn{1}{|c|}{P8}  & \multicolumn{1}{c|}{\checkmark \{VC1, VC2, VC3\}}                          & \checkmark \{VC5\}                     & \checkmark \{VC1, VC2, VC3, VC4\}                                          & \multicolumn{1}{c|}{\checkmark \{VC1, VC2, VC3\}}                                                                  & \checkmark \{VC2\}                                                                                                          & \checkmark \{VC2\}                                                                                    \\ \hline
\multicolumn{1}{|c|}{P9}  & \multicolumn{1}{c|}{\checkmark \{VC1, VC2\}}                               & \checkmark \{VC2, VC3, VC5\}           & \checkmark \{VC4\}                                                         & \multicolumn{1}{c|}{\checkmark \{VC1, VC2, VC3\}}                                                                  & $\times$                                                                                                                    & $\times$                                                                                              \\ \hline
\multicolumn{1}{|c|}{P10} & \multicolumn{1}{c|}{\checkmark \{VC1, VC2, VC3, VC5\}}                     & \checkmark \{VC1, VC2, VC3, VC5\}      & \checkmark \{VC1, VC2, VC4\}                                               & \multicolumn{1}{c|}{\checkmark \{VC1, VC2, VC3, VC5\}}                                                             & \checkmark \{VC2\}                                                                                                          & \checkmark \{VC2\}                                                                                    \\ \hline
\multicolumn{1}{|c|}{P11} & \multicolumn{1}{c|}{\checkmark \{VC1\}}                                    & $\times$                               & \checkmark \{VC4\}                                                         & \multicolumn{1}{c|}{\checkmark \{VC1, VC2\}}                                                                       & $\times$                                                                                                                    & \checkmark \{VC2\}                                                                                    \\ \hline
\end{tabular}
}
\end{table*}

\subsection{Observation and Results}

As shown in \Cref{tab:qual_study_1}, 6 participants failed to answer \textbf{T3(Q2)} and 5 failed to answer \textbf{T4(Q1)}, thereby indicating that performing tasks \textbf{T3} and \textbf{T4} was difficult for our participants using this prototype.  Also, for \textbf{T1(Q2)}, 3 participants failed to answer correctly, and 1 participant could not answer \textbf{T2(Q1)}. However, all our participants could successfully answer \textbf{T1(Q1)} and \textbf{T3(Q1)}. 

\begin{figure*}
  \centering
  \includegraphics[width=0.9\linewidth]{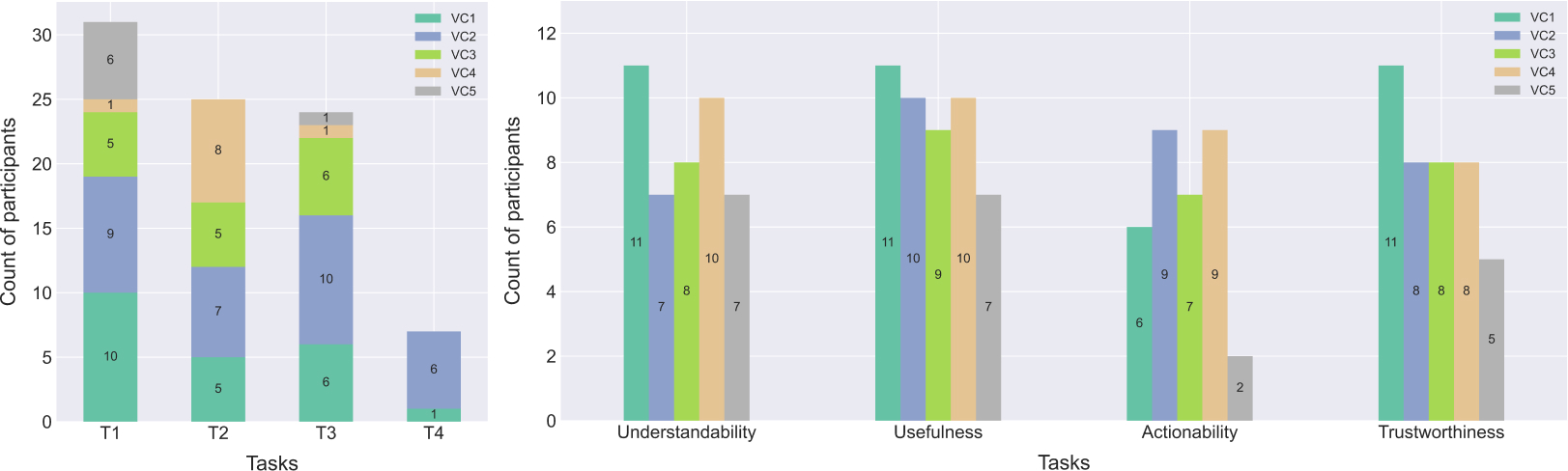}
  \caption{Results from our qualitative evaluation of our low-fidelity prototype. (left) Chart showing the count of participants using the visual components for the task-based questions. (right) Comparison of each visual component in terms of understandability, usefulness, actionability and trustworthiness by reported by the participants.}
  \Description[Results from our qualitative evaluation of our low-fidelity prototype]{Results from our qualitative evaluation of our low-fidelity prototype}
  \label{fig:study1_analysis}
\end{figure*}

\Cref{fig:study1_analysis} shows the preferred visuals for each task and compares all the visual components based on understandability, usefulness, actionability, and trust based on the participant's responses. Although each visual was designed to support a specific set of tasks, it was observed that for all the tasks, there is a high preference for the \textit{Patient Summary} (\textbf{VC2}) as the majority found it more informative because of the graphical representation of the data with clear color-coding. More than half of the participants considered all the visual components useful. But the textual patient information with the risk chart presented in \textbf{VC1} is considered the most useful, followed by the graphical representation of patient data in \textbf{VC2} and recommendations in \textbf{VC4}. In terms of actionability, \textbf{VC2} and \textbf{VC4} were considered the most actionable. \textbf{VC1} was also considered the most trustworthy, while only 5 participants considered the risk recovery information in \textbf{VC5} as trustworthy as the others either found it difficult to interpret or did not consider it important. 

We relate our observations with the following themes generated using our thematic analysis process.

\emph{Visualizations facilitate quick understanding}: All the participants were positive about the entire dashboard but the visual representation of the patient records enabled them to understand the patient's condition quickly. Most of them connected the usefulness and understandability of the visual components with graphical and color-coded representations. For example, P10 stated, “\textit{It's clearly indicated that the risk is high. I think it's very colorfully indicated and that's good because it looks like it's marked in colors from green, orange, and red}”. This justifies why the \textit{Patient Summary} (\textbf{VC2}) was highly preferred by our participants.

\emph{Including textual annotations can improve understandability}: Although the \textit{Patient Summary} \textbf{VC2} was highly preferred, some participants expressed difficulty in interpreting the data-distribution charts in \textbf{VC2}: “\textit{I’m not sure about the distribution [in patient summary] and what does that tell … what does the peak mean?}”(P9). This justifies why many participants failed to answer \textbf{T4(Q1)} correctly. They preferred more textual representation provided in \textbf{VC1} and \textbf{VC3}. However, it is important to keep the balance between the amount of textual and graphical representations, as too much textual information can affect the usefulness of the visual: “\textit{I think it is very important [to have graphical representations] … not just numbers, as it should tell you what will happen with the numbers. It makes it likable and useful and easy to understand}” (P6). While a lack of concise textual description can also impact the interpretation of the visual, like in the case of \textit{Risk Recovery} (\textbf{VC5}), almost half of our participants, found it difficult to interpret “\textit{I don’t know what it [risk recovery] means. I need more text and information on this}” (P2). 

\emph{Interactive visuals increase the interpretability of explanation methods}: Most of our participants liked how they could interact with the dashboard to observe changes in the predicted risk: “\textit{Using it [interactions] you can show them [patients] if they have lower blood sugar, then what will happen. So, you are not just telling them, they can see how it improves their health}” (P6), “\textit{I think that when you see that way if you reduce the blood sugar, how the graph is changing, it would be a motivation from them [patients]}” (P5). Interactions allow them to explore and easily understand visual explanations. This indicates that exploration through interactive visuals can increase the interpretability of the method. On the contrary, less interactive visual like the \textit{Risk Recovery} (\textbf{VC5}) was difficult to understand for our participants “\textit{I don’t understand the risk recovery too well. That can be improved}” (P3). As the interpretation of \textbf{VC5} was not very clear to many, they failed to answer \textbf{T1(Q2)} correctly. However, we observed that the discoverability of interactions was very difficult using the click-through prototype. This justifies why many participants failed to answer \textbf{T3(Q2)}.

\emph{Combination of visual components}: It was observed that most participants combined two or more visuals to perform the given tasks. Particularly, we observed them combining multiple visuals when suggesting actions (\textbf{T2}). For interpreting the logic behind the predictions and drawing comparisons with other patients they mostly used the patient summary (\textbf{VC2}). Some of the participants mentioned all the visuals were useful, and it was hard for them to comment if a visual was more useful than others. Thus, they considered the entire dashboard very useful, actionable, and trustworthy. P10 stated: “\textit{It’s hard to say [which one is better] because I think all of them are very good. This [VC1] provides the basic information you need to know first. This [VC2] is like a really good summary because it shows you in detail what's going on with the patient … Well, here's the risk recovery, and you can see in the graph that risk is elevating. From risk factors, I can figure out what increased risk. This [VC4] recommends how to reduce risks. So, I would say all of them combined have their own role}”.  

\emph{Association of trust with visualization and underlying data}: None of the participants mentioned a lack of trust due to “complex algorithms”. They could trust the system as they could see the reference patient data: “\textit{Yeah, I trust this not because it's generated by the computer, but the overall score is evidence-based [based on patient data]}” (P9). When asked about why the system is predicting the risk of diabetes as high, all of them mentioned values of the health variables which were higher than the recommended ranges and mentioned the color-coding of red used to denote health factors that are not good for the patient: “\textit{[Risk is high] because you see the high-risk circle and a lot of things are in color red, and we also see that the blood sugar is 7.5, much higher than the recommended range. This is the biggest indicator of diabetes}” (P11). So, their sense of trust is linked with the visual representation of the patient data. Moreover, lack of interpretation of the visuals (like for \textbf{VC5}) affects the overall trust and hence the usefulness of the visual.  Like P2 and P3, even P11 did not consider the risk recovery visual trustworthy as they did not understand it: “\textit{I really don't know. It's something to do with months, so with time and the risk. What I don't know is what it should mean}” (P11).

\emph{Action recommendation based on data, risk predictions, and a priori knowledge}: For suggesting actions to reduce the risk of diabetes, most participants relied on the reference data and their interpretation of how the underlying patient data is related to the overall risk. Their ability to perform interactions with the patient data to observe changes in the predicted risk helped them to suggest actions: “\textit{if the blood sugar is lower say 5.8, the overall risk is in orange and I see that risk is lower and so moderate}” (P6). However, most of them used their a priori knowledge of the domain to suggest actions to minimize risk: “\textit{The highest risk is [due to] the blood sugar as we can see in the chart because the red level [from VC2] is the biggest and I would recommend a diet for the blood sugar}” (P5). Even feature-importance-based \textbf{VC3} and counterfactual-based recommendations provided in \textbf{VC4} were less preferred for suggesting actions as they expressed the need to have additional information about the patient's diet, current medications, blood pressure, and other information not used by the ML model: “\textit{Something about the patient’s diet is missing in this dashboard … nutrition is not included and whether the patient is taking a specific medication for diabetes or other reasons}” (P4). However, we observed a higher preference for data-centric explanations provided through \textbf{VC2} for action recommendation compared to counterfactual explanations provided through \textbf{VC4} as they considered the recommendations to be very generic and not personalized: “\textit{It [VC4] is useful. But it's like a general suggestion for everyone. I think they [patients] trust much more when we don't generalize them with another. This kind of recommendation would be too general for one patient}” (P5).

\emph{Patients as potential users}: All our participants mentioned that our dashboard can be a useful tool for monitoring patient health conditions: “\textit{I think it will be great tool for the healthcare providers, because you can't remember everything. And if you have all the data combined in one platform, it will be very good and you can see the progress, the risk factors all at once. So, you don't have to put them together from the different lab results}” (P8). Additionally, most of them considered the dashboard to be a good source of motivation for patients to create a better awareness of their health conditions: “\textit{Patients also need to see some graphs to realize that their health is not so good}”(P3). However, they expressed some concern for older patients to directly use this dashboard: “\textit{The younger patients would use this but for the older patients I don't know}” (P8). Even though our prototype was designed around the needs of HCPs, an interesting point of analyzing the patient’s perspective was raised from their feedback. Thus, we included patients as participants along with HCPs during the evaluation of our high-fidelity prototype.

\begin{figure*}
  \centering
  \frame{\includegraphics[width=0.90\linewidth]{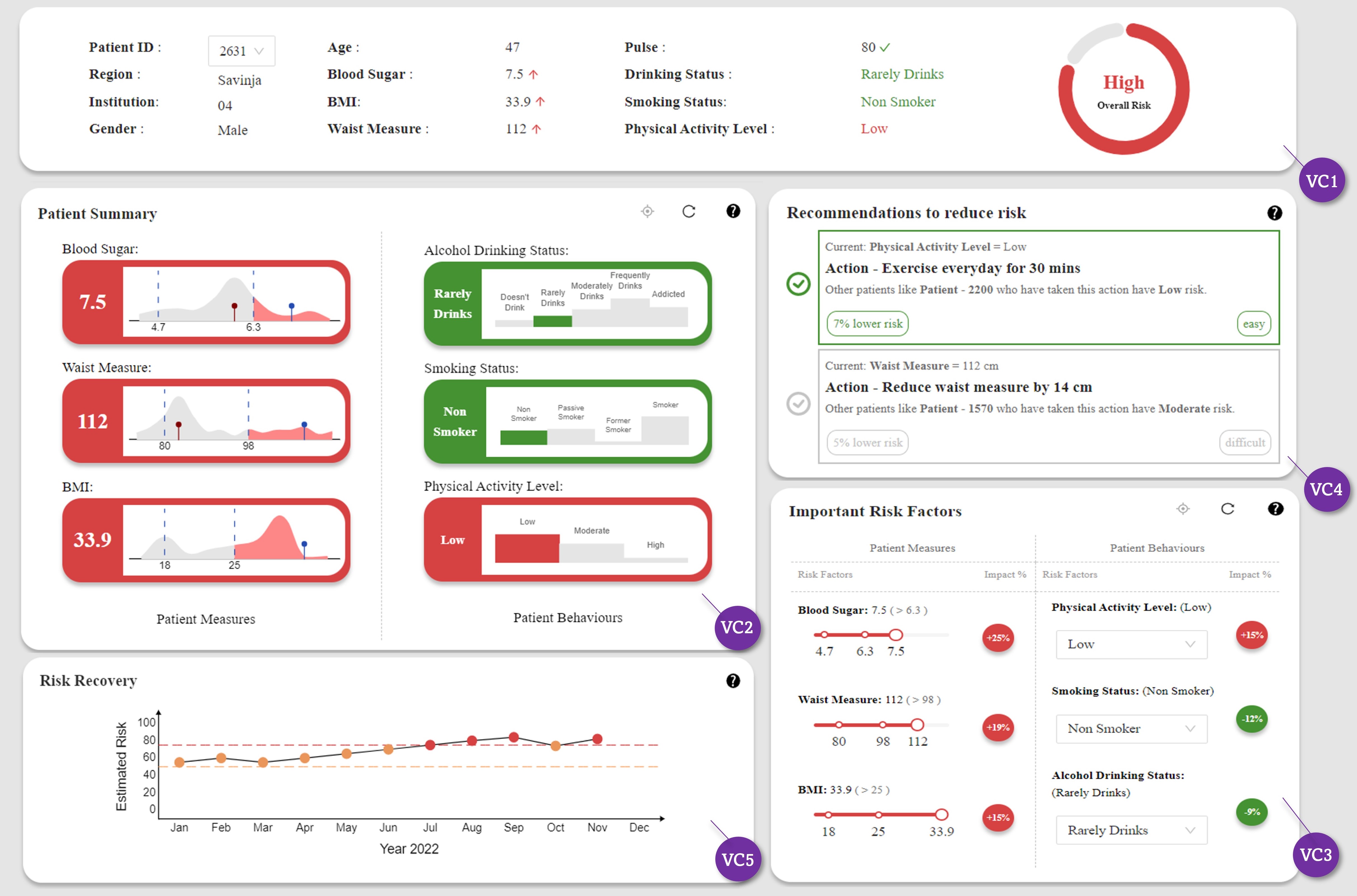}}
  \caption{Dashboard design of the high-fidelity web application prototype. Visual explanations are provided using: Patient information with the risk prediction chart (VC1), Patient Summary (VC2), Important Risk Factors (VC3), Recommendations to reduce risk (VC4), Risk Recovery (VC5). This is the modified version of the dashboard used for final evaluation.}
  \Description[Dashboard design of the high-fidelity web application prototype]{Dashboard design of the high-fidelity web application prototype}
  \label{fig:hi_fi}
\end{figure*}

\section{Evaluation and Analysis of High-fidelity Prototype}
After analyzing the qualitative data collected from the previous study, we implemented a high-fidelity web application prototype developed using Meteor, React.js, HTML, and CSS. The various charts used in our interactive dashboard are developed using Chart.js and the application is deployed using Docker.  \cref{fig:hi_fi} presents a snapshot of our high-fidelity prototype. The prototype was evaluated through an online user study in Prolific \cite{45} as our recruitment platform and Qualtrics \cite{46} as our survey platform. 

\subsection{Design Rationale}
From our first study, we observed that some of our participants found it difficult to discover the interactive components from our click-through prototype. Thus, we added icons as explicit indicators for the visuals which supported \textit{what-if interactions} in our high-fidelity prototype. Additionally, we added tooltips with short descriptions for each visual component. Furthermore, the mouse cursor style was modified when the user hovered over the interactive components for encouraging them to click and observe the changes in the interactive plots. 

As many participants found the \textit{Risk Recovery} (\textbf{VC5}) difficult to interpret, we added interactive effects on hover that highlights the risk zones of the patient. We aimed to increase user exploration through interactions to increase understandability of this visual.

We also observed that our participants in our first study found the \textit{Factors contributing to risk} (\textbf{VC3}) to be less actionable than the \textit{patient summary} (\textbf{VC2}) and \textit{Recommendations to reduce risk} visual (\textbf{VC4}). Some of our participants suggested adding what-if interactions to \textbf{VC3} for making it more actionable. Thus, we have considered this feedback and modified \textbf{VC3} for supporting what-if interactions. Another feedback was received for simplifying the title of \textbf{VC3} and so we renamed it \textit{Important Risk Factors}. We also swapped the position of \textbf{VC3} and \textbf{VC4} in the high-fidelity prototype to improve the discoverability of the recommendations as suggested by our participants.

Hence, we considered the feedback received through our observation and participant responses during the evaluation of our click-through prototype to improve the design of our high-fidelity prototype to effectively meet our user requirements.

\begin{table*}
\caption{Information of participants recruited for the evaluation of the high-fidelity prototype}
\begin{subtable}[h]{.45\linewidth}
\centering
\caption{HCPs}
\label{tab:participants_hcp_study2}
 \scalebox{.75}{
\begin{tabular}{ll}
\toprule
& \textbf{Participant distribution}                               \\ \midrule
Country & \begin{tabular}[c]{@{}l@{}}
United Kingdom (12/45)\\ 
South Africa (8/45)\\ 
Mexico (5/45)\\ 
Poland (4/45)\\ 
United States, Portugal (3/45)\\ 
Chile (2/45)\\ 
Finland, Sweden, Hungary, \\
Israel, Canada, Spain, \\
Italy, Germany (1/45)
\end{tabular}                    
\\ \midrule
Gender & \begin{tabular}[c]{@{}l@{}}
17 : male\\ 
27 : female \\ 
1 : non-binary
\end{tabular}                                                     \\ \midrule
Age group & \begin{tabular}[c]{@{}l@{}}
26 : (21 - 30) years\\ 
11 : (31 - 40) years\\ 
2 : (41 - 50) years\\ 
6 : (51 - 60) years
\end{tabular}                                                                        \\ \midrule
Highest education level & \begin{tabular}[c]{@{}l@{}}
1: Ph.D.\\ 
9 : Master's degree\\ 
35 : Bachelor's degree
\end{tabular}                                                                        \\ \midrule
Experience in patient care  & \begin{tabular}[c]{@{}l@{}}
12 : < 1-year\\ 
16 : (1 - 3) years\\ 
3 : (3 - 5) years\\ 
1 : (5 - 10) years\\ 
13 : > 10-years
\end{tabular} \\ 
\bottomrule
\end{tabular}}
\end{subtable} %
\hfill
\begin{subtable}[h]{.45\linewidth}
\centering
\caption{Patients}
\label{tab:participants_patient_study2}
 \scalebox{.75}{
\begin{tabular}{ll}
\toprule
& \textbf{Participant distribution}                                                 
\\ \midrule
Country & \begin{tabular}[c]{@{}l@{}}
United States (17/51)\\ 
South Africa (13/51)\\ 
United Kingdom (8/51)\\ 
Portugal (4/51)\\ 
Spain (3/51)\\ 
Poland (2/51)\\ 
Czech Republic, Estonia, \\
Norway, Italy (1/51)
\end{tabular}                    
\\ \midrule
Gender & \begin{tabular}[c]{@{}l@{}}
27 : male\\ 
24 : female
\end{tabular}                                                     \\ \midrule
Age group & \begin{tabular}[c]{@{}l@{}}
1 : < 20-years\\ 
10 : (21 - 30) years\\ 
8 : (31 - 40) years\\ 
9 : (41 - 50) years\\ 
13 : (51 - 60) years\\ 
10 : > 60-years
\end{tabular}                                                     \\ \midrule
Highest education level & \begin{tabular}[c]{@{}l@{}}
5 : Master's degree\\ 
24 : Bachelor's degree\\ 
21 : High School degree\\ 
1 : Preferred not to disclose
\end{tabular}                                                     \\ \midrule
Diabetes duration  & \begin{tabular}[c]{@{}l@{}}
2 : < 1-year\\ 
7 : (1 - 3) years\\ 
11 : (3 - 5) years\\ 
7 : (5 - 10) years\\ 
24 : > 10-years
\end{tabular} \\ 
\bottomrule
\end{tabular}}
\end{subtable} 
\end{table*}

\begin{table*}[h]
\caption{Grading rubric considered during evaluation of the high-fidelity prototype.}
\label{tab:grading_rubric}
 \scalebox{.85}{
\begin{tabular}{ll}
\toprule
\textbf{Correct with sufficient justification}   & Correct response with all correct rules and no extra, unnecessary rules            \\ 
\midrule
\textbf{Correct with insufficient justification} & Correct response but with only some rules or extra, unnecessary rules              \\ 
\midrule
\textbf{Guess/unintelligible}                    & Correct response but with no reason, or with wrong interpretation of the visual(s) \\ 
\midrule
\textbf{Incorrect }                              & Failed to give correct responses                          \\ 
\bottomrule
\end{tabular}
}
\end{table*}

\subsection{Participants}

This prototype was evaluated using 45 HCPs like nurses, physicians, and medical workers and 51 diabetic patients. We recruited patients who had been diagnosed with diabetes at least 6 months prior to our experiment and HCPs' who were actively involved in patient care. Using Prolific, the recruited participants were compensated with an hourly rate of \$10 for their time. 

\Cref{tab:participants_hcp_study2} presents the demographic information of our HCP participants. Collectively, they had experience in dealing with all types of patients in any age group with both acute and chronic disorders (not just diabetes) from non-critical to critical nature.  The demographic information of our patient participants is presented in  \cref{tab:participants_patient_study2}. All of our patient participants were actively in contact with their HCPs and they had awareness of the risk factors of type-2 diabetes.

\subsection{Procedure}

We first gave an overview of the prototype to our participants and suggested them to explore it on their own. Then, they were given similar task-based questions as the previous study based on the four supported tasks (\textbf{T1, T2, T3, T4}) through an online questionnaire. Based on the information shown in our prototypical dashboard, our participants were asked to identify patients with the most critical condition, their current risk level, and whether their condition is improving or not for task \textbf{T1}. For \textbf{T2}, they were asked to suggest actions to reduce the high risk for a specific patient shown on our dashboard. For \textbf{T3}, they were asked to justify why the system indicated that a specific patient had a high risk while another had a low risk. Finally, for \textbf{T4}, they were asked to compare a specific patient's health factors with the recommended range of values for the health factor and with those of the other patients. 

Additionally, our participants had to justify their responses to the task-based questions. We also asked them to report their perception of the visual components in terms of understandability, usefulness and trustworthiness through 4-point Likert Scale questions. We included additional open-ended questions to gather more qualitative data about actionability of our dashboard and their motivation for using it. 

During the evaluation process, we categorized all the responses to the task-based questions into four categories: \textit{correct with sufficient justification}, \textit{correct with insufficient justification}, \textit{guess / unintelligible} and \textit{incorrect}, similar to Lim et al. \cite{48} as shown in \Cref{tab:grading_rubric}. We recorded their overall response time and mouse-movements to track their interactions with our dashboard while answering the given questions. Furthermore, we analyzed the qualitative responses about the actionability of our dashboard to understand why the participants found it actionable and which component is considered most actionable. Additionally, we categorized the qualitative responses about motivation to use our dashboard as \textit{positive}, \textit{negative} or \textit{neutral} to analyze their sentiments and understand the rationale behind their responses.

We performed hypothesis testing with one-proportion z-test at 5\% significance level \cite{44} to measure the statistical significance of the correct responses to the task-based questions. We aimed to achieve to a success rate of 80\% for the tasks supported by our dashboard. Thus, our null hypothesis ($H_{0}$) was that 80\% of the participants giving correct responses, while our alternate hypothesis ($H_{A}$) is more than 80\% giving correct responses. We further noted the proportion of participants giving correct responses with sufficient and insufficient justifications. We used descriptive statistics for the evaluation of the remaining questions considering their format instead of hypothesis testing.

\begin{figure*}
  \centering\includegraphics[width=0.90\linewidth]{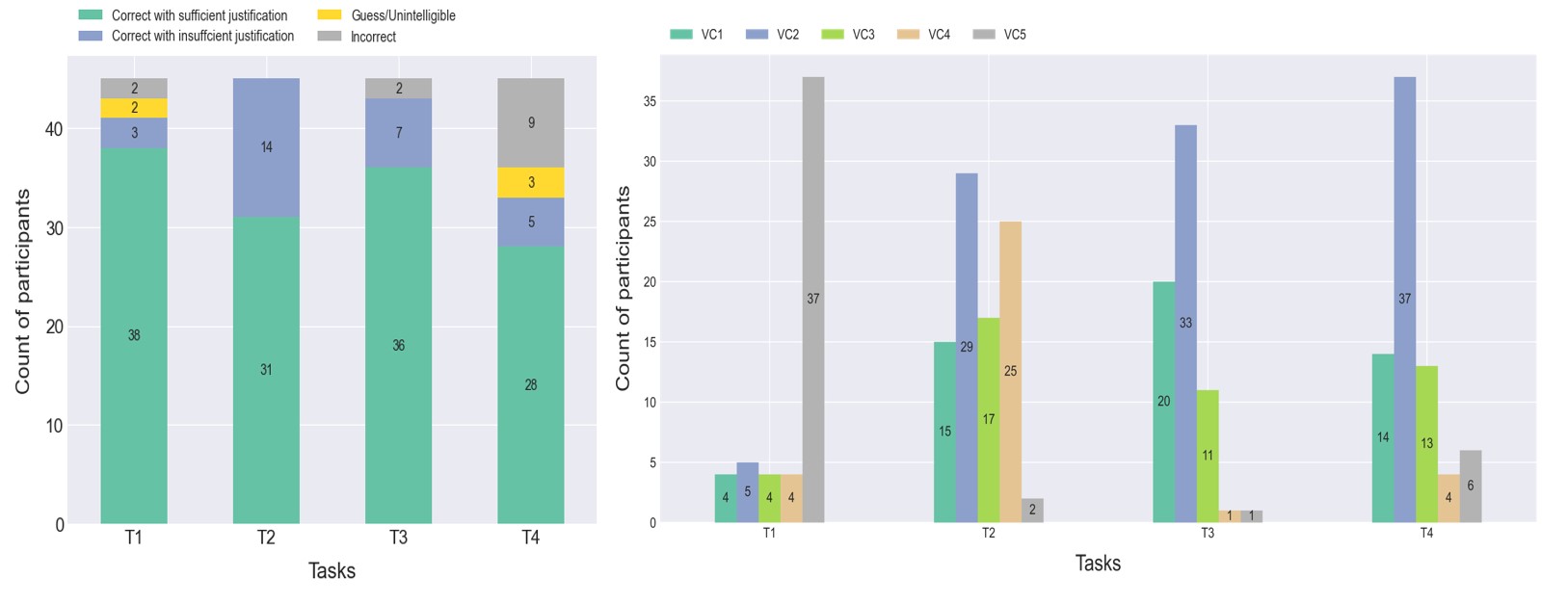}
  \caption{Results obtained from the responses of the HCP participants. (left) Responses are categorized according to the grading rubric in \Cref{tab:grading_rubric} for the tasks supported. (right) Chart showing the count of the participants using the visual components to answer the task-based questions.}
  \Description[Results obtained from the responses of the HCP participants]{Results obtained from the responses of the HCP participants}
  \label{fig:study2_hcp}
\end{figure*}

\begin{figure*}
  \centering
  \includegraphics[width=1.00\linewidth]{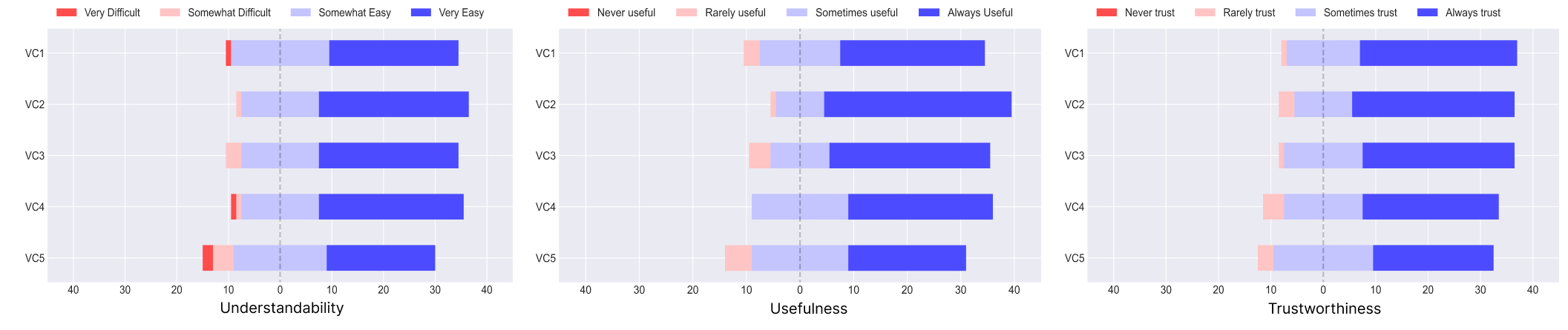}
  \caption{Diverging bar-charts summarizing the self-reported Likert-scale responses of the HCP group}
  \Description[Self-reported Likert-scale responses of the HCP group]{Diverging bar-charts summarizing the self-reported Likert-scale responses of the HCP group}
  \label{fig:uat_hcp}
\end{figure*}

\subsection{Observation and Results}

\emph{HCP participants}: We observed that 41 HCPs (91.1\%) gave correct responses for \textbf{T1} questions (\textit{z =2.619, p= 0.0044}). It was observed that 84.4\% of them provided sufficient justifications, while only 6.67\% HCPs failed to provide sufficient justifications to their correct responses. They mostly used the risk recovery visual to perform this task. 
For questions asked for task \textbf{T2}, all 45 HCPs gave correct responses (\textit{z=$\infty$, p=0.00}). However, 31.1\% of them failed to provide sufficient justifications. 
For task \textbf{T3}, 43 HCPs (95.56\%) gave correct responses (\textit{z=5.064, p=0.00}). But only 4.4\% of the correct responses did not include sufficient justifications. 
Although most of them reported using the \textit{patient summary} (VC2) to perform tasks \textbf{T3} and \textbf{T4}, we observed them combining multiple visuals like VC2, VC3 and VC4. This indicates that the participants preferred combining different explanation methods, i.e. data-centric, feature-importance, and counterfactual explanations when suggesting actions and understanding the rationale behind the predicted risk. For \textbf{T4}, 33 HCPs (73.3\%) gave correct responses (\textit{z=-1.011, p=0.8441}), out of which 11.1\% did not include sufficient justifications. They mostly used VC2 to perform this task.

We only \textit{failed to reject $H_{0}$} for task \textbf{T4}, even though we observed a high proportion (73.3\%) of the HCP participants giving correct responses. For other tasks, we \textit{rejected $H_{0}$}, suggesting that more than 80\% of HCPs can perform these tasks correctly. On investigating the reason why 9 HCPs (20\%) gave incorrect answers to T4 questions, we observed that unlike the majority of the HCPs who used the \textit{patient summary} (\textbf{VC2}), they used the patient information provided in \textbf{VC1} to manually compare different patient records by using the patient id filter. This involved many interactions and it was not convenient to perform the comparisons manually. Therefore, many of them failed to give the correct responses. \cref{fig:study2_hcp} illustrates visual components used by HCPs for all the task-based questions and gives an overview of their responses to all the tasks.

\begin{figure*}
  \centering\includegraphics[width=0.90\linewidth]{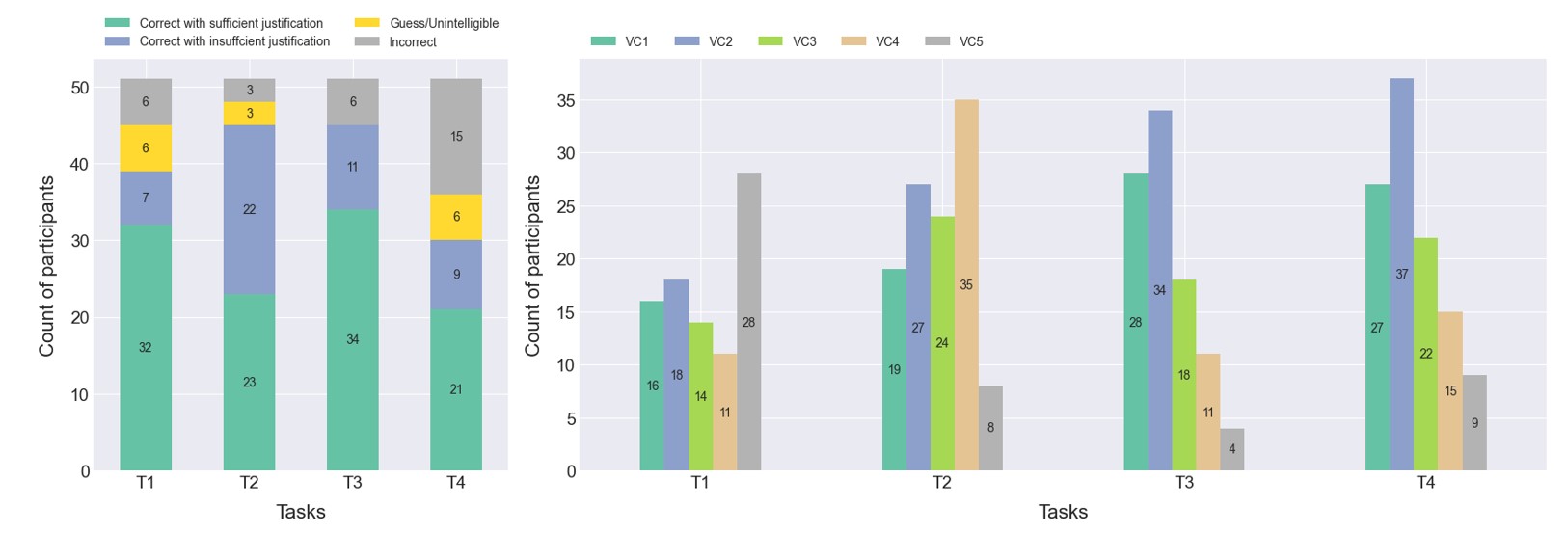}
  \caption{Results obtained from the responses of the patient participants. (left) Responses are categorized according to the grading rubric in \Cref{tab:grading_rubric} for the tasks supported. (right) Chart showing the count of the participants using the visual components to answer the task-based questions.}
    \Description[Results obtained from the responses of the patient participants]{Results obtained from the responses of the patient participants}
  \label{fig:study2_patient}
\end{figure*}

\begin{figure*}
  \centering
  \includegraphics[width=1.00\linewidth]{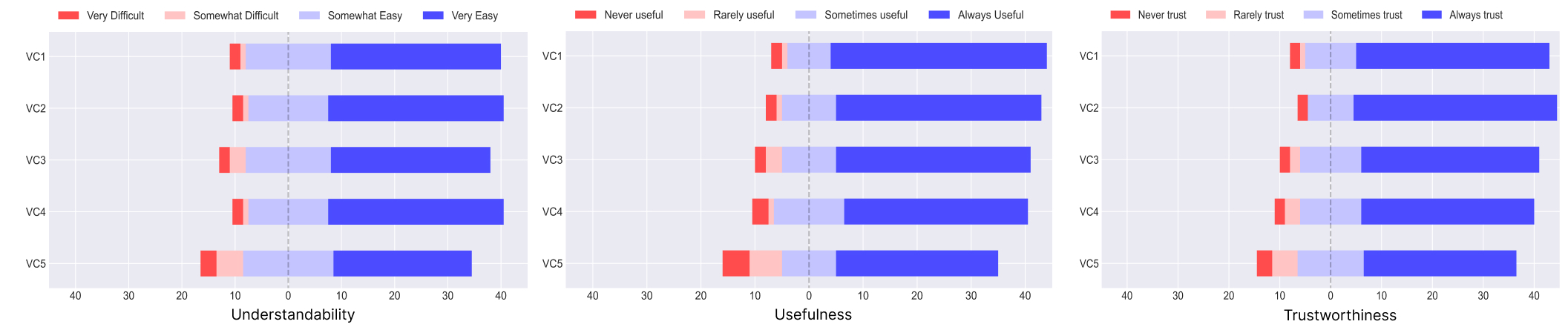}
  \caption{Diverging bar-charts summarizing the self-reported Likert-scale responses of the patient group}
  \Description[self-reported Likert-scale responses of the patient group]{Diverging bar-charts summarizing the self-reported Likert-scale responses of the patient group}
  \label{fig:uat_patients}
\end{figure*}

\cref{fig:uat_hcp} illustrates the self-reported responses to the Likert-scale questions about understandability, usefulness, and trustworthiness of the visual components of the HCP group. This group found the \textit{Patient Summary} (\textbf{VC2}) the easiest to understand and the most useful. For trustworthiness, the HCPs found \textbf{VC1} to be most trustworthy. \Cref{tab:likert_responses} presents the proportion of participants for each self-reported Likert scale question for both groups.

\begin{table*}[h]
\caption{Results showing the proportion of participants for the self reported responses to 4-point Likert Scale questions about Understandability, Usefulness, Trustworthiness}
\label{tab:likert_responses}
 \scalebox{.7}{
\begin{tabular}{@{}lcllllllllll@{}}
\toprule
&    & \multicolumn{2}{c}{\textbf{VC1}} & \multicolumn{2}{c}{\textbf{VC2}} & \multicolumn{2}{c}{\textbf{VC3}} & \multicolumn{2}{c}{\textbf{VC4}} & \multicolumn{2}{c}{\textbf{VC5}} \\
\midrule
&    & \multicolumn{1}{c}{HCPs} & Patients & HCPs     & Patients     & HCPS     & Patients     & HCPS     & Patients     & HCPs     & Patients     \\
\cmidrule(lr){3-4}
\cmidrule(lr){5-6}
\cmidrule(lr){7-8}
\cmidrule(lr){9-10}
\cmidrule(lr){11-12}
\multirow{4}{*}{\textbf{Understandability}}
& Very Easy to Understand & 55.5\%  & 62.7\% & 64.4\% & 64.7\% & 60\% &  58.8\%  & 62.2\% &  64.7\% & 46.7\% & 50.9\%             
\\
& Somewhat Easy to Understand & 42.2\%  & 31.3\% & 33.3\% & 29.4\% & 33.3\% &  31.3\%  & 33.3\% &  29.4\% & 40\% & 33.3\%             
\\
& Somewhat Difficult to Understand & 0\%  & 1.9\% & 2.2\% & 1.9\% & 6.6\% &  5.8\%  & 2.2\% &  1.9\% & 8.8\% & 9.8\%             
\\
& Very Difficult to Understand & 2.2\%  & 3.9\% & 0\% & 3.9\% & 0\% &  3.9\%  & 2.2\% &  3.9\% & 4.4\% & 5.8\%             
\\
\midrule
\multirow{4}{*}{\textbf{Usefulness}} 
& Always Useful & 60\%  & 78.4\% & 77.7\% & 74.5\% & 66.6\% &  70.5\%  & 60\% &  66.6\% & 48.8\% & 58.8\%             
\\
& Sometimes Useful & 33.3\%  & 15.6\% & 20\% & 19.6\% & 24.4\% &  19.6\%  & 40\% &  25.4\% & 40\% & 19.6\%             
\\
& Rarely Useful & 6.6\%  & 1.9\% & 2.2\% & 1.9\% & 8.8\% &  5.8\%  & 0\% &  1.9\% & 11.1\% & 11.7\%             
\\
& Never Useful & 0\%  & 3.9\% & 0\% & 3.9\% & 0\% &  3.9\%  & 0\% &  5.8\% & 0\% & 9.8\%             
\\
\midrule
\multirow{4}{*}{\textbf{Trustworthiness}} 
& Always Trust & 66.6\%  & 74.5\% & 68.8\% & 78.4\% & 64.4\% &  68.6\%  & 57.7\% &  66.6\% & 51.1\% & 58.8\%             
\\
& Sometimes Trust & 31.1\%  & 19.6\% & 24.4\% & 17.6\% & 33.3\% &  23.5\%  & 33.3\% &  23.5\% & 42.2\% & 25.4\%             
\\
& Rarely Trust & 2.2\%  & 1.9\% & 6.6\% & 0\% & 2.2\% &  3.9\%  & 8.8\% &  5.8\% & 6.6\% & 9.8\%             
\\
& Never Trust & 0\%  & 3.9\% & 0\% & 3.9\% & 0\% &  3.9\%  & 0\% &  3.9\% & 0\% & 5.8\%             
\\
\bottomrule

\end{tabular}}
\end{table*}

The majority (64.4\%) of the HCPs considered \textbf{VC2} as the most actionable. They mentioned that the color-coded and graphical representation of the health variables helped them to easily identify risk factors that need immediate attention for suggesting actions. However, most of them mentioned using two or more visuals together for suggesting actions and understanding the logic behind the predicted risk. For instance, one of the participants mentioned: “\textit{Given the patient summary it would make it easier to address the most pressing risk factors for a patient and what they need to work on at most}”.

For the motivation of using our dashboard, 44 HCPs (97.7\%) responded positively for using it as a screening tool during their consultation with patients. They mentioned about using our dashboard to show the risk levels and create more awareness for the patients.

\emph{Patient participants} : We observed that 39 patients (76.5\%) gave correct responses to \textbf{T1} questions (\textit{z=-0.594, p=0.724}). Despite 62.7\% of them giving sufficient justification for their correct responses, we \textit{failed to reject} $H_{0}$. 
But as 45 of them (88.2\%) could answer \textbf{T2} questions correctly (\textit{z=1.825, p=0.034}), we could \textit{reject} $H_{0}$. However, we observed 43.14\% of them failed to give sufficient justification for their correct responses. 
On investigating further, we observed that 58\% of the older patients (>50 years) failed to give sufficient explanations for \textbf{T2} questions, indicating their understanding of the visual components used to recommend actions was not fully correct. Like \textbf{T2} questions, 45 of them (88.2\%) could answer \textbf{T3} questions correctly (\textit{z=1.825, p=0.034}). So, we could \textit{reject} $H_{0}$ for \textbf{T3} questions. 
But we observed 66.67\% of them giving sufficient justifications, while 21.57\% gave insufficient justification for their correct responses. For \textbf{T4}, we observed only 33 (64.7\%) could give correct responses (\textit{z=-2.286, p=0.98}). Despite 54.9\% of them giving sufficient justification for their correct responses, we \textit{failed to reject} $H_{0}$. 
Additionally, we observed that most older patients (>50 years) struggled with \textbf{T4} as 50\% of the older patients failed to give sufficient explanations for \textbf{T4} questions.  \cref{fig:study2_patient} illustrates visual components used by the patients for all the task-based questions and gives an overview of their responses to all the tasks.

\Cref{fig:uat_patients} illustrates the responses of the patient group to the Likert-scale questions about understandability, usefulness and trustworthiness of the visual components. This group found both \textbf{VC2} and \textbf{VC3}  as the easiest to understand as compared to other visuals. However, they found \textbf{VC1} to be the most useful and \textbf{VC2} to be most trustworthy.

We observed the majority (68.6\%) of the patients also considered \textbf{VC2} as the most actionable as they could easily identify the high-risk factors similar to the HCP group. Like the HCPs, even this group had mentioned combining multiple visuals together for all the given tasks. For instance, one of the patient participants mentioned: “\textit{The ‘Recommendations to reduce risk gives precise ways to reduce my risk of diabetes. The ‘Patient Summary’ section showed me areas where my behavior, such as physical exercise, was too low in certain areas}”. 

We observed that 48 patients (94.11\%) responded positively when asked about their motivation to use this dashboard.  Most of them mentioned interactive explanations increased their motivation as they could see how the risk changes on changing the values of the health variables.

\section{Discussion}
\subsection{Key Takeaways From Our User Studies}

From our first user study, we collected feedback on the design of our visually directive explanation methods from HCPs. We used their feedback to make design changes discussed in Section 5.1 for our web application prototype. We also collected an early indication of the usefulness, understandability, actionability and trustworthiness of the visual components from our participants.  

In our second study, we collected data from a larger pool of participants to validate our observations from the previous study. The web application used in our second study was interactive and enabled our participants to give better feedback. Also, the feedback collected from patients along with HCPs helped us to justify the effectiveness of our explanation methods included in our dashboard.

Results from our second study indicate that a significant proportion of HCPs could successfully perform all the given tasks. Most of them could provide sufficient justifications for their correct responses irrespective of their age group, domain expertise, or years of experience in patient care. However, we observed older patients (> 50 years) struggled with tasks for suggesting actions to reduce risk (\textbf{T2}) and comparing their health measures with that of other patients (\textbf{T4}). Further simplifications may be needed for this user group.

Overall, our participants used data-centric explanation based \textit{patient summary} (\textbf{VC2}) and \textit{patient information} (\textbf{VC1}) more than other visuals for all the tasks. While explaining their responses, most of them compared the current health measures with the recommended values of health variables. Color-coded representations were considered more useful than graphical representations of the data. This indicates that color-coded and interactive data-centric explanations form a vital part of our explanation dashboard for both HCPs and patients. 

 Unlike the HCPs, we observed a high proportion of patients (68.6\%) using the counterfactual-based recommendations (\textbf{VC4}) for suggesting actions to minimize risk (\textbf{T2}). Some of them mentioned that for getting quick suggestions, they preferred the recommendation list. But to get more detail on how these recommendations are helpful, they relied on data-centric explanations. 
 
 Moreover, both HCPs and patients have mentioned combining multiple visual components, especially for suggesting actions to minimize high risk and understanding the rationale behind the predicted risk. This suggests that the limitations of any explanation method included in our dashboard can be complemented by the other methods.

\subsection{Addressing the Research Questions}
\textbf{RQ1. In what ways do patients and HCPs find our visually directive explanation dashboard useful for monitoring and evaluating the risk of diabetes onset?} – Our results indicate that both HCPs and patients found our explanation dashboard very useful for monitoring and evaluating the risk of diabetes onset. As inferred from our user studies, the interactive visual components enabled our participants to explore and unravel the rationale behind the predicted risk. 
They mentioned that interactive and directive visuals, that show how the predicted risk changes when changing the health variables, are useful to create more awareness for the patients and drive change in their behavior.

\textbf{RQ2. In what ways do HCP and patients perceive data-centric, model-centric, and example-based visually directive explanations in terms of usefulness, understandability, and trustworthiness in the context of healthcare?} – It was observed that most of our participants justified the rationale behind the predicted risk by referring to the underlying patient data used for training the ML model and its color-coded visual representations. They mentioned about trusting the predictions as they could easily relate the prediction with the underlying data. Thus, we get an indication that data-centric explanations are more trustworthy and useful than commonly adopted model-centric feature-importance explanations, especially in healthcare. 

However, as they mentioned using a combination of visual components together, the significance of feature-importance explanations cannot be neglected. Additionally, our participants have mentioned that the what-if interactions enabled them to explore our dashboard and develop a better understanding of the visual explanations. Also, our participants found the example-based counterfactual explanations important and useful when they wanted explanations in the form of recommendations. 

Furthermore, our participants have shown a higher usage of our representation of data-centric explanations through the \textit{patient summary} (\textbf{VC2}) for performing given tasks and actions over other explanation methods as they found them more informative. However, in general, it was observed that more visual, interactive explanations with reference to the recommended range of data having concise textual descriptions are more useful to both HCPs and patients.

\textbf{RQ3. In what ways do visually directive explanations facilitate patients and HCPs to take action for improving patient conditions?} – Patients reported that interactive explanations increased their motivation of using our dashboard as a self-screening tool as they could see how the risk changes on changing the health variables. While HCPs wanted to use this dashboard for better communication with patients during their consultations. They found our dashboard to be actionable and useful as they can utilize it for guiding patients in improving their health by showing the high-risk factors. From the qualitative data collected through our user studies, we inferred that the interactive visual explanations enabled both HCPs and patients to explore how to alter the predicted outcome along with explaining the factors that could affect the predictions.

\subsection{Tailoring Directive Explanations for Healthcare Experts}
We share our design implications for tailoring the visual representation of directive explanations for healthcare experts from our observations and results. Our design implications are aligned with the recommendations from Wang et al.'s framework \cite{35}. 

\emph{Increasing actionability through interactive what-if analysis}: During the evaluation of our low-fidelity prototype, our participants highlighted that the conventional representation of feature-importance explanations (as illustrated by \textit{factors contributing to risk} visual in \Cref{fig:low_fi}) was less actionable than data-centric explanations presented through the \textit{patient summary} \textbf{(VC2)} as it was difficult for them to understand how the risk factors affected the predicted risk. Our modified design of this visual component (\textbf{VC3}) used in our high-fidelity prototype enabled them to perform interactive what-if analysis, i.e. allowed them to change the feature values and observe the change in the overall prediction. Hence, we recommend the usage of interactive design elements that allows what-if analysis for representing directive explanations for HCPs. This recommendation also \textit{supports hypothesis generation} as proposed by Wang et al. \cite{35}

\emph{Explanations through actionable features instead of non-actionable features}: In our approach, we included only \textit{actionable variables} for visual components which supports what-if interactions and better \textit{identification of coherent factors} \cite{35}. We anticipated that allowing the ability to alter values of non-actionable variables can create confusion for HCPs, especially for representing counterfactual explanations. Thus, we propose providing explanations through actionable variables for suggesting actions that the user can perform to obtain their desired outcomes.

\emph{Color-coded visual indicators}:  HCPs indicated that the color-coded representations of risk factors were very useful for getting quick insights. Hence, we recommend the usage of color-coded representations and visual indicators to highlight factors that can increase or decrease the predictor variable. This suggestion further facilitates Wang et al.'s recommendation \cite{35} of \textit{identifying coherent factors}.

\emph{Data-centric directive explanations}: HCPs indicated that our representation of data-centric explainability through the patient summary was very informative. They could easily identify how good or bad the risk factors are for a specific patient. Additionally, they could get an overview of how other patients are doing as compared to a specific patient through the data-distribution charts. Thus, our representation of data-centric explainability provided a local explanation but with a global perspective. This suggestion is also aligned with the recommendations from Wang et al. \cite{35} as data-centric directive explanations support \textit{forward reasoning} by providing \textit{access to source and situational data} and yet can be \textit{easily integrated with multiple explanation methods}.

\subsection{Limitations and Future Work}
In this section, we articulated some limitations of this work: (1) Our prototype used offline predictions about the overall risk generated by our model instead of real-time predictions. The use of other ML algorithms with real-time inference processes might impact the perceived utility of the tool. (2) The prototype was evaluated with HCPs with backgrounds in diverse specializations. Even though we were able to reach a wider population, it would be more helpful to evaluate this prototype with HCPs who are dedicated to taking care of patients with type-2 diabetes. (3) The importance of different explanation methods was examined jointly as part of the dashboard and not analyzed independently. Consequently, the limitations of some of the explanation methods could be concealed by the benefits of other methods. (4) Since the prototype was personalized for monitoring the risk of diabetes onset, the findings from this research may not be applicable for monitoring other diseases as the user needs for other diseases can be very distinct. 

 In our future studies, we aim to focus on personalizing directive counterfactual explanations, as our participants had expressed a need for a better representation of such explanations. Additionally, we plan to analyze the utility of different explanation methods used in the dashboard in isolation. 

\section{Conclusion}
In this research work, we present a directive explanation dashboard that combines visually represented data-centric, feature-importance, and counterfactual explanations for monitoring the risk of diabetes onset. Our research compared the different visual explanations in terms of understandability, usefulness, actionability, and trustworthiness with healthcare experts and patients. Our participants have shown a higher preference for the visually represented data-centric explanations that provided local explanations with a global overview, over other methods. Especially, we observed that the color-coded risk factors and data-distribution charts in our visually directive data-centric explanations assisted healthcare experts in suggesting actions to reduce risk by easily identifying high-risk factors. Based on our results, we suggest using such data-centric explanations combined with other explanations. We hope our results will inspire other researchers to utilize such visually directive explanations to other specialization areas in healthcare as well. 

\begin{acks}
We would like to thank Oscar Alvarado, Robin De Croon, Maxwell Szymanski, Houda Lamqaddam and Diego Rojo for providing helpful comments that improved this text. We also thank Lucija Gosak for helping us with participant recruitment for our first user study. This work was supported by Research Foundation–Flanders (FWO, grant G0A3319N) and KU Leuven Internal Funds (grant C14/21/072).
\end{acks}

\bibliographystyle{ACM-Reference-Format}
\bibliography{references}

\end{document}